\documentclass[aps,ac,12pt,superscriptaddress,notitlepage]{revtex4-2}
\usepackage[utf8]{inputenc}
\usepackage{amsmath, amssymb, bm}       
\usepackage{graphicx}                   
\usepackage{wrapfig}                    
\usepackage{array}                      
\usepackage{booktabs}                   
\usepackage{setspace}                   
\usepackage{placeins}                   
\usepackage{chngcntr}                   
\usepackage{braket}
\usepackage{hyperref}
\UseRawInputEncoding
\usepackage{xr}                         

\usepackage{tikz}                       
\usetikzlibrary{trees}                  

\begin{document}
\title{Nonreciprocal magneto-optic beam steering of thermal emission}

\author{Mason Christiansen}\affiliation{Department of Materials Science and Engineering, University of California, Los Angeles, Los Angeles, CA 90095, USA}
\author{Daniel Cui}\affiliation{Department of Materials Science and Engineering, University of California, Los Angeles, Los Angeles, CA 90095, USA}
\author{Aaswath P. Raman}
\email{aaswath@ucla.edu}
\affiliation{Department of Materials Science and Engineering, University of California, Los Angeles, Los Angeles, CA 90095, USA}
\affiliation{California NanoSystems Institute, University of California, Los Angeles, Los Angeles, CA 90095, USA}

\begin{abstract}
Conventional thermal emitters obey Kirchhoff's law of thermal radiation, which ties absorptivity to emissivity and places a fundamental limit on a range of radiative energy technologies and sensing systems. This constraint can be overcome using magneto-optic materials, which exhibit nonreciprocal thermal emission under an applied magnetic field. While the spectral nonreciprocity of magneto-optic devices has been explored theoretically and experimentally, nonreciprocal control over the direction of thermal emission has received far less attention. Here, we propose and numerically demonstrate nonreciprocal beam steering of thermal emission, wherein an applied magnetic field lifts the degeneracy between guided modes at $\pm k_x$, enabling asymmetric and tunable directional thermal emission. We highlight a III-V semiconductor photonic structure operating at $24.5\,\mu$m, a wavelength with few laser sources. This structure exhibits a nonreciprocity greater than $0.5$ between $\theta=\pm 75^\circ$ at $50\,$mT, a peak emissivity asymmetry over 0.6 at $150\,$mT, and $22^\circ$ of beam steering between $0$ and $1\,$T. Our approach provides a solid-state means of steering a thermal beam, offering a potential route to magnetically controlled beam scanning of incandescent sources.
\end{abstract}

\maketitle

\section{Introduction}

Thermal emission is a fundamental form of electromagnetic radiation central to many natural and technological processes. Controlling the properties of thermal emission and absorption is of interest for many technologies due to its applications in energy conversion \cite{Sakr_2017,Davids_2020}, imaging \cite{Hulley_2016,Lucchi_2018,vollmer_2021}, sensing \cite{Palchetti_2016}, and radiative cooling \cite{Raman_2014,Suhendri_2020}. A large body of work has been dedicated to engineering the spectral and directional emissivity of materials by utilizing surface plasmon \cite{Wang_2017}, surface phonon polariton \cite{Greffet_2002} and epsilon-near-zero (ENZ) modes \cite{Hwang_2023,Sakr_2017}, as well as through photonic crystal \cite{Lin_2003} and metamaterial-based architectures \cite{Zhang_2026, Liu_Tyler_2011}. Additionally, the ability to dynamically modulate the spectral and directional emissivity of a material has also been explored \cite{Wang_2017,Liu_2011,Vassant_2013,Park_2018,Dong_2019, Hwang_2025}. However, these techniques, using reciprocal materials and devices, are constrained by Kirchhoff's law of thermal radiation.

Kirchhoff's law states that for a particular wavelength of light, angle from the surface normal of the material, and polarization, the absorptivity and emissivity of the material are equal. This places fundamental limits on the efficiency of a range of radiative energy technologies as well as sensing systems~\cite{Ghalekohneh_Nonrecip_Solar_TPV2022,park_reaching_2022,Park_2022}. However, Kirchhoff's law can be violated in materials that violate Lorentz reciprocity, including materials exhibiting a strong magneto-optic response, making it possible for emissivity and absorptivity to differ at a given wavelength and angle of incidence. A range of structures have been proposed on this basis, including magneto-optic photonic crystals \cite{Zhu_2014}, gratings and guided-mode resonant structures \cite{Zhao_2019,Zhao_2021}, and magnetic Weyl semimetals, which break reciprocity without any external field \cite{Pajovic_2020,Kotov_2018}. Recent work has also experimentally demonstrated nonreciprocal thermal emission primarily by applying large magnetic fields ($>$ 1 T) to heavily doped III-V semiconductor-based systems \cite{Shayegan_2023,Shayegan_2022,Nabavi_2025,Zhang_2025,Liu_2023}. In nearly all of this work the goal is to decouple the absorption and emission channels at a given wavelength, since doing so might improve the efficiency of solar and thermophotovoltaic energy conversion technologies \cite{Green_2012,Buddhiraju_2018,Ghalekohneh_Nonrecip_Solar_TPV2022,park_reaching_2022,Park_2022}.

By contrast, leveraging nonreciprocity to control the direction of thermal emission has remained largely unexplored. Breaking time-reversal symmetry does not simply decouple absorptivity and emissivity at a fixed angle; it can also lift the degeneracy between modes propagating with $+k_x$ and $-k_x$. In a reciprocal system, Lorentz reciprocity requires the scattering matrix to be symmetric \cite{Jalas_2013}, which constrains the dispersion relation to satisfy $\omega(k_x) = \omega(-k_x)$. A magnetic perturbation breaks this symmetry, displacing the dispersion in opposite directions for opposite wavevectors. Because the emission angle is set by $k_x$, this displacement is a change in the direction of peak emission that is controlled entirely by an external magnetic field. Nonreciprocity therefore offers not only spectral control but a route to magnetically steering a thermal beam, without moving parts and without the symmetric $\pm\theta$ emission that constrains conventional directional emitters. It is this angular, rather than spectral, consequence of nonreciprocity that we pursue here.

In this work, we propose and numerically demonstrate a magneto-optic approach to beam steering that enables nonreciprocal and asymmetric control of thermal emission. In particular, we present an all-III-V semiconductor structure that exhibits a nonreciprocal contrast greater than $0.5$ between $\theta=\pm 75^\circ$ at $50$ mT, a peak emissivity asymmetry over 0.6 at $150$ mT, and $22^\circ$ of beam steering between $0$ and $1$ T for $24.5$ $\mu$m light. Unlike conventional thermal emission, which is isotropic and broadband, the emission from this structure is confined in both direction and wavelength and is strongly asymmetric between $\pm\theta$. The device operates in the 20--25 $\mu$m range, where few laser sources exist \cite{Wilson_2019}. Furthermore, modulating the applied magnetic field offers a potential route to beam scanning with an incandescent source.

\begin{figure}[htbp]
  \centering
  \includegraphics[width=6.47in]{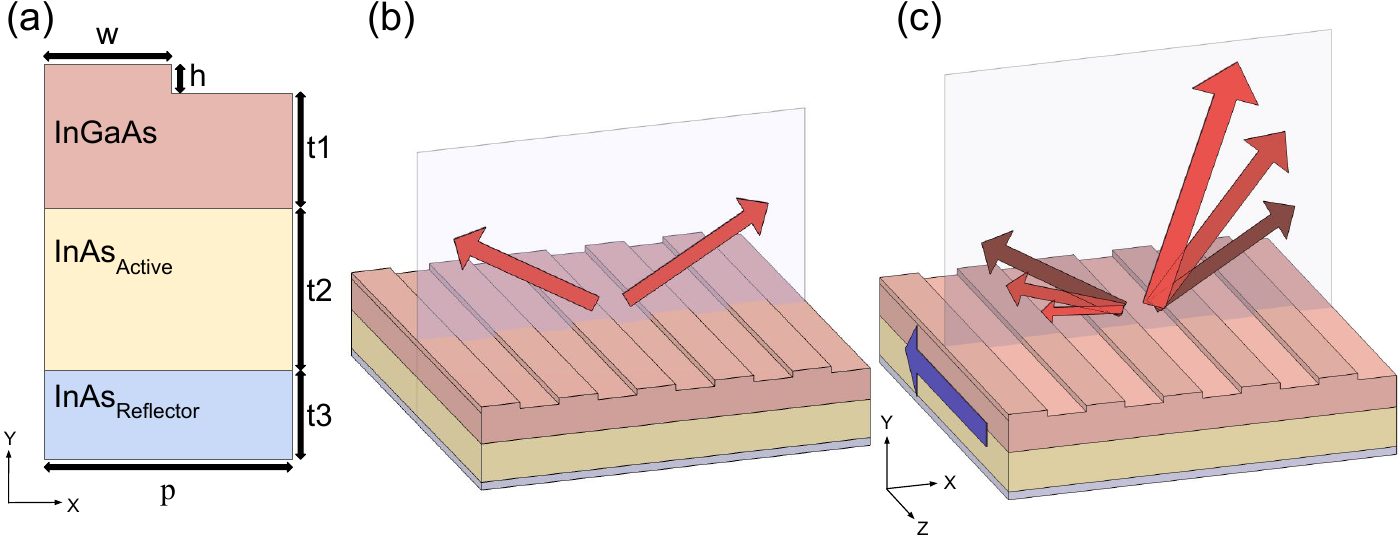}
\caption{(a) Schematic of one period of the device. The bottom layer consists of degenerately doped InAs and serves as an optical reflector, while the middle layer consists of lower doped InAs and serves as the active layer. The top layer is intrinsic In$_{0.53}$Ga$_{0.47}$As with a grating on the surface. (b) A 3D rendering of the device showing the symmetric emissivity with zero magnetic field. (c) The change in direction and magnitude of emissivity with the application of a magnetic field in the direction of the blue arrow.}
\label{fig1}
\end{figure}

\section{Results}

As our model system, we consider a structure composed of three layers of III-V semiconductors, as shown in Fig.~\ref{fig1}(a). The top layer has a thickness, $t_1$, of 4$\,\mu$m (not including the grating), is intrinsic In$_{0.53}$Ga$_{0.47}$As, and has a grating etched into its surface which supplies the in-plane momentum needed to couple free-space radiation into the guided modes of the structure. The middle layer is doped InAs, has a thickness, $t_2$, of 5$\,\mu$m, and is the active layer responsible for the nonreciprocal behavior of the device. The bottom layer is degenerately doped InAs, has a thickness, $t_3$, of 1$\,\mu$m, and acts as a reflector in this wavelength range due to its high carrier density. The grating has width, $w$, of 5$\,\mu$m, height, $h$, of 0.27$\,\mu$m, and the structure has a period, $p$, of 10$\,\mu$m. The carrier concentration, effective electron mass, and scattering rate of each layer are summarized in Table~\ref{tbl1}.

The doping concentration of the active layer sets the plasma frequency and thereby determines the ENZ wavelength. A concentration of $7.8\times10^{17}\,$cm$^{-3}$ places the ENZ condition at $24.1\,\mu$m. The active layer's $5\,\mu$m thickness is sufficient to support a confined surface mode while remaining thin enough that the mode retains appreciable field amplitude at the upper interface. The degenerately doped bottom layer has a plasma wavelength of $1.8 \,\mu$m and is therefore metallic across the entire band considered here, so it acts as a back reflector that suppresses transmission. The intrinsic In$_{0.53}$Ga$_{0.47}$As top layer is transparent in this range and its $4\,\mu$m thickness positions the antinode of the resonance near the active layer.

The nonreciprocal behavior arises from the magneto-optic response of the active layer, where an applied magnetic field introduces off-diagonal components in the permittivity tensor
\begin{align}
\varepsilon =
\begin{pmatrix}
\varepsilon_{xx} & \varepsilon_{xy} & 0 \\
\varepsilon_{yx} & \varepsilon_{yy} & 0 \\
0 & 0 & \varepsilon_{zz}
\end{pmatrix}
\end{align}
whose elements are defined as
\begin{align}
\varepsilon_{xx} &= \varepsilon_{yy}
= \varepsilon_\infty - \frac{\omega_p^2(\omega + i\gamma)}
{\omega\left[(\omega+i\gamma)^2-\omega_c^2\right]} \\
\varepsilon_{xy} &= -\varepsilon_{yx}
= i \frac{\omega_p^2\omega_c}
{\omega\left[(\omega+i\gamma)^2-\omega_c^2\right]} \\
\varepsilon_{zz} &= \varepsilon_\infty
- \frac{\omega_p^2}{\omega(\omega+i\gamma)}.
\end{align}
Here $\gamma$ is the scattering rate, $\omega_p=\sqrt{n e^2/(m_e \varepsilon_0)}$ is the plasma frequency, $\omega_c = e B/m_e$ is the cyclotron frequency, $n$ is the free carrier concentration, and $B$ is the applied magnetic field. When $B=0$, $\varepsilon_{xx}=\varepsilon_{yy}=\varepsilon_{zz}$ and the off-diagonal components of the permittivity tensor, $\varepsilon_{xy}$ and $\varepsilon_{yx}$, are zero, so the material is isotropic and the dispersion satisfies $\omega(k)=\omega(-k)$. When a magnetic field is applied the permittivity tensor acquires antisymmetric off-diagonal components as the free carriers move in cyclotron motion causing $\varepsilon_{xy}$ and $\varepsilon_{yx}$ to become non-zero, so the material no longer obeys time-reversal symmetry. The dispersion now has a directional dependence such that $\omega(k) \neq \omega(-k)$ so $\alpha(\theta) \neq \alpha(-\theta)$, and because $\epsilon(\theta)=\alpha(-\theta)$ we can conclude that $\epsilon(\theta) \neq \epsilon(-\theta)$.

Throughout, we take $\hat{z}$ to lie parallel to the grating grooves, $\hat{x}$ along the direction of periodicity in the plane of the surface, and $\hat{y}$ normal to the surface. The applied magnetic field is oriented along $\hat{z}$, parallel to the grooves and perpendicular to the plane of incidence, which is the Voigt configuration. In this geometry the off-diagonal elements $\varepsilon_{xy}$ and $\varepsilon_{yx}$ couple the
in-plane and out-of-plane field components of transverse-magnetic polarized light, so all emissivity reported here is for TM polarization. Transverse-electric polarized light has its electric field along $\hat{z}$ and therefore samples only $\varepsilon_{zz}$, which is unaffected by the magnetic field; the TE emissivity is symmetric in $\theta$ at all fields and is not considered
further in our analysis.

\begin{figure}[htbp]
  \centering
  \includegraphics[width=6.47in]{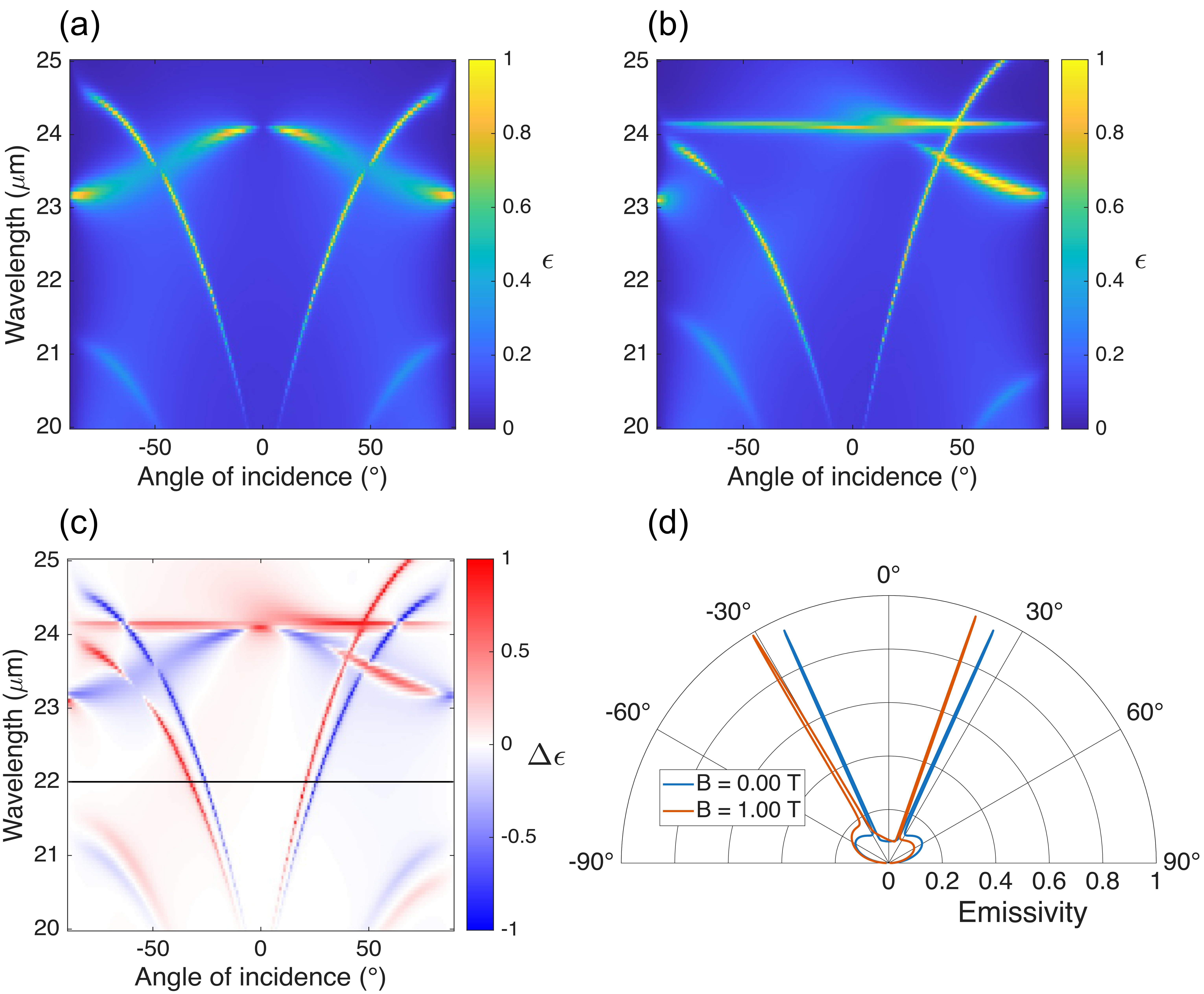}
\caption{The spectral and directional emissivity of the device at (a) $0\,$T, and (b) $1\,$T. (c) The difference between the spectral and directional emissivity at 1 and $0\,$T. (d) The emissivity as a function of angle at $\lambda=22\,\mu$m for 1 and $0\,$T (shown by the black line in (c)).}
\label{fig2}
\end{figure}

For this structure, at wavelengths longer than the ENZ wavelength of the active layer, as $B$ increases the emissivity for $+\theta$ grows in magnitude and is shifted towards the normal, while the emissivity for $-\theta$ shrinks in magnitude and goes towards oblique angles, as shown qualitatively in Fig.~\ref{fig1}(b) and (c). Fig.~\ref{fig2} shows the simulated spectral and directional emissivity of the device for $B=0\,\mathrm{T}$ and $B=1\,\mathrm{T}$. As expected, in Fig.~\ref{fig2}(a) the emissivity is perfectly symmetric across the normal because there is no applied magnetic field so there is no nonreciprocity. That same symmetry is broken in Fig.~\ref{fig2}(b) when a $1\,$T field is applied.

In addition to the nonreciprocal mode, the emissivity becomes nearly
omnidirectional at the ENZ wavelength of $24.1\,\mu$m.
This is a consequence of the flat dispersion of the ENZ mode: as
$\varepsilon_{xx} \rightarrow 0$ the resonant frequency becomes nearly
independent of $k_x$, so a broad range of in-plane wavevectors, and
therefore a broad range of emission angles, satisfies the resonance
condition at essentially the same wavelength.

The mode of interest in Fig.~\ref{fig2}(a) and (b) is the guided-mode resonance (GMR) that originates near normal incidence at $20\,\mu$m and shifts toward more oblique angles with increasing wavelength. The magnetic field makes the GMR dispersion nonreciprocal, causing the resonances at $+k_x$ and $-k_x$ to shift differently. Because the GMR follows a dispersion relation, $\omega(k_x)$, a magnetic-field-induced change in resonant frequency is accompanied by a change in the corresponding in-plane wavevector, which produces a shift in the angle of peak emission, resulting in beam steering. The largest beam steering occurs where the GMR dispersion is relatively flat. In this regime, a small magnetically induced shift in the resonance wavelength produces a large change in the emission angle.

Fig.~\ref{fig2}(c) shows the difference between Fig.~\ref{fig2}(a) and (b) and most clearly visualizes beam steering for the entire mode. The blue curve shows where the mode is at zero applied field and the red curve shows where the mode is shifted when $1\,$T is applied. For a given wavelength, the angular separation from the blue mode to the red mode (on the same side of the normal) is the amount of beam steering done. Fig.~\ref{fig2}(d) shows beam steering for a particular wavelength, $22 \, \mu$m, as shown by the black line in Fig.~\ref{fig2}(c). Notice that because the magnetic field moves the mode to longer wavelengths for positive angles and toward shorter wavelengths for negative angles, the $+\theta$ peak moves towards the normal while the $-\theta$ peak moves away from the normal.

\begin{figure}[htbp]
  \centering
  \includegraphics[width=6.47in]{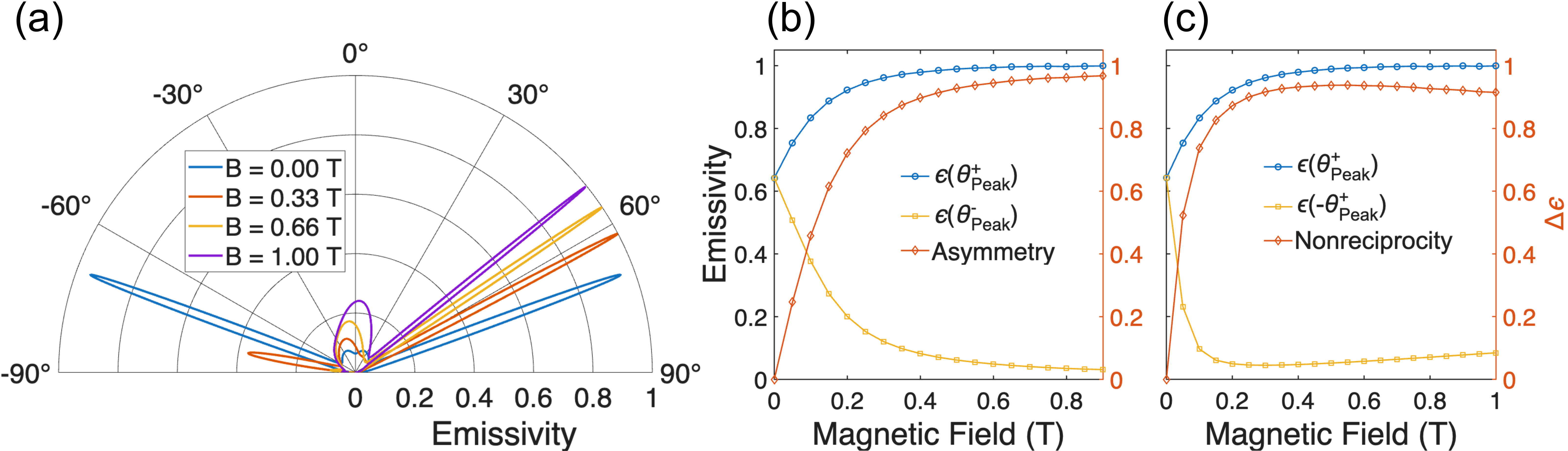}
\caption{(a) The emissivity of the device as a function of angle at 24.5 $\mu$m for multiple magnetic field values. (b) Asymmetry as a function of magnetic field, defined as the difference between the peak emissivity at $+\theta$ and the tracked peak emissivity at $-\theta$. (c) Nonreciprocity as a function of magnetic field, defined as the difference between the peak emissivity at $+\theta$ and the emissivity at the corresponding symmetric angle $-\theta$. The respective emissivities are also shown. }
\label{fig3}
\end{figure}

While beam steering is observed at wavelengths near $22\,\mu$m, the emissivity peaks at $+\theta$ and $-\theta$ remain approximately equal. In contrast, for wavelengths greater than the ENZ wavelength of the active layer, the structure exhibits both significant beam steering and strong asymmetry between the emissivity peaks.

Figure~\ref{fig3}(a) shows the angular emissivity at $24.5\,\mu$m for several applied magnetic fields. As the magnetic field increases, the peak emission for $+\theta$ shifts toward the normal and grows in magnitude while the corresponding peak at $-\theta$ decreases substantially in magnitude and eventually disappears. This asymmetry is quantified as the difference between the peaks at $+\theta$ and $-\theta$, as shown in Fig.~\ref{fig3}(b), and exceeds 0.6 at $150\,$mT. At sufficiently high magnetic fields, a separate near-normal emission peak becomes larger than the $-\theta$ peak. This near-normal feature, which arises from a weak Fabry-Perot resonance [Fig.~\ref{fig4}(c)], is excluded from the analysis. We track only the pair of peaks present at zero magnetic field and their evolution in angle and magnitude with increasing magnetic field.

There is also significant nonreciprocity at this wavelength. We define nonreciprocity as the difference between the peak emissivity at $+\theta$ and the emissivity at the corresponding negative angle, as shown in Fig.~\ref{fig3}(c). Because $\epsilon(-\theta)=\alpha(\theta)$, this is equivalent to $\epsilon(\theta)-\alpha(\theta)$, and therefore directly quantifies the violation of Kirchhoff's law. Notably, the nonreciprocity is greater than 0.5 with an applied field of only $50\,$mT.

\section{Modeling \& Discussion}
\begin{figure}[htbp]
  \centering
  \includegraphics[width=6.47in]{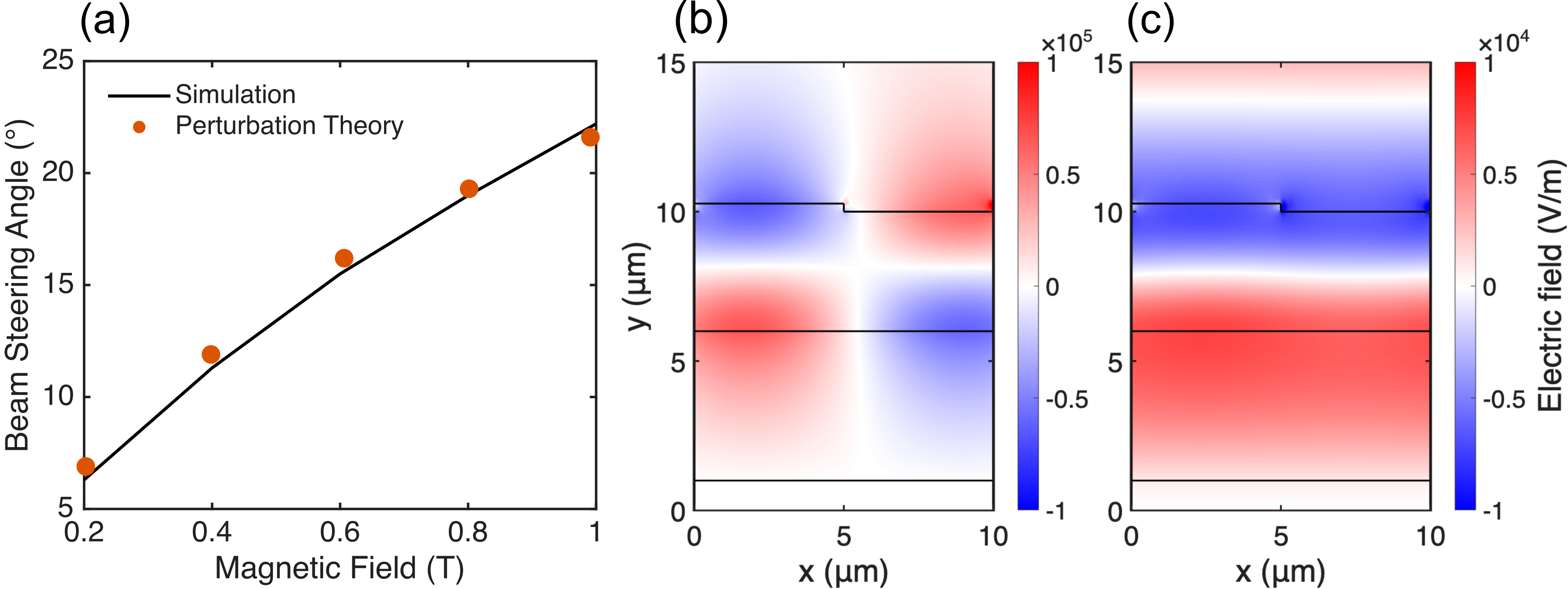}
\caption{(a) Beam steering angle as a function of magnetic field, as predicted by simulation and perturbation theory. (b) The $x$-component of the electric field at $\lambda=24.5\,\mu$m, $B=1\,$T, and $\theta=54.1^\circ$, corresponding to the angle of maximum emissivity. (c) The $x$-component of the electric field at $\lambda=24.5\,\mu$m, $B=1\,$T, and $\theta=0^\circ$.}
\label{fig4}
\end{figure}

The amount of beam steering is shown in Fig.~\ref{fig4}(a). For an applied field of $1\,$T, the peak emission is steered by $22^\circ$. Furthermore, the simulated steering angle agrees well with that predicted by the perturbation theory developed in Ref.~\cite{Cui_2026}, providing confidence that the observed steering originates from the magnetically induced modification of the mode dispersion. The perturbation theory predicts the shift in resonance frequency from the applied magnetic field. The inputs are the in-plane electric field components and out-of-plane magnetic field component of the unperturbed mode, along with the relevant material properties. The resulting frequency shift is
\begin{equation}
\Delta \omega = -\omega_c F,
\label{eq_dw}
\end{equation}
where
\begin{equation}
F =
\frac{
\dfrac{\omega_p^2 \varepsilon_\infty}{\omega^2}
\displaystyle\iint_{\mathrm{active}}
\operatorname{Im}
\left(
E_x^* E_y - E_y^* E_x
\right)
\,dx\,dy
}{
\displaystyle\iint
\left\{
\frac{1}{2}
\left[
\varepsilon_\infty
\left(
|E_x|^2 + |E_y|^2
\right)
+
\mu_0 |H_z|^2
\right]
+
\frac{1}{2\varepsilon_\infty\omega_p^2}
\left(
|V_x|^2 + |V_y|^2
\right)
\right\}
\,dx\,dy
}.
\label{eq_F}
\end{equation}
Here $\mathbf{V}=\partial\mathbf{P}/\partial t$ is the time derivative of the free-carrier polarization field $\mathbf{P}$, the integral in the numerator is taken over the magneto-optic active layer, and the denominator is the sum of the kinetic and potential energies of the unperturbed mode \cite{Cui_2026}.

To apply this perturbation theory to beam steering at a fixed wavelength, we invert Eq.~\eqref{eq_dw} using $\omega_c = eB/m_e$ to determine the magnetic field required to produce a specified frequency shift:
\begin{equation}
B = -\frac{m_e\,\Delta \omega}{e F}.
\label{eq_B(dw)}
\end{equation}

We then use the zero-field dispersion of the GMR to relate a desired change in emission angle, $\Delta\theta$, to the corresponding wavelength shift, $\Delta\lambda$, required to move the resonance to the target wavelength of $24.5\,\mu$m. This $\Delta\lambda$ is converted to a frequency shift, $\Delta\omega$, which is then used in Eq.~\eqref{eq_B(dw)} to calculate the required magnetic field. Repeating this procedure for a range of steering angles yields the relationship between applied magnetic field and beam steering angle, which is compared with the full-wave simulation in Fig.~\ref{fig4}(a).

The $x$-component of the electric field, $E_x$, at the angle of maximum emissivity for $1\,$T, shown in Fig.~\ref{fig4}(b), is consistent with a guided-mode resonance. In contrast, $E_x$ at normal incidence, Fig.~\ref{fig4}(c), is characteristic of a weak Fabry-Perot resonance which is responsible for the near-normal emissivity growing with increasing magnetic field.

Because the practical utility of beam steering depends on the steering range relative to the width of the beam itself, we also extract the angular full width at half maximum of the emissivity peak. At zero field the peak at
$24.5\,\mu$m has a FWHM of $5.6^\circ$, and at $1\,$T the steered peak narrows to $1.9^\circ$ as the mode shifts toward the normal. The $22^\circ$ steering range is therefore nearly four times the zero-field FWHM. We note that the linewidth is set by the total loss rate of the mode, with contributions from the free-carrier scattering rate $\gamma$ of the active layer and from radiative leakage through the grating.

Prior experimental demonstrations of nonreciprocal thermal emission have relied on fields exceeding $1\,$T, which in practice requires a superconducting or large electromagnet and effectively confines the effect to the laboratory~\cite{Shayegan_2023,Shayegan_2022}. By contrast, the structure considered here produces an emissivity difference greater than $0.5$ at $50\,$mT, a
field readily produced by a small permanent magnet or a compact air-core coil. Modulation and beam steering can in principle be achieved with a small electromagnet assembly. Because the steering mechanism is magnetic rather than mechanical, the modulation rate is set by the electromagnet rather than by the emitter. For a coil of inductance $L$ and resistance $R$ the field follows the drive current with a time constant $L/R$, and for a compact coil producing $50\,$mT this corresponds to a bandwidth on the order of 100 Hz. Finally, we note that the results above are computed at room temperature. Raising the temperature increases the phonon scattering rate and therefore $\gamma$, broadening the resonance and potentially reducing its peak emissivity. Additionally, any temperature dependence of the carrier concentration may shift the ENZ wavelength, necessitating appropriate design modifications that take this into account.

\section{Conclusion}
We have introduced an all-III-V semiconductor structure and shown through simulation that it is capable of generating a greater than $50\%$ difference in asymmetric emissivity between $\theta=\pm 75^\circ$ at $50\,$mT and $22^\circ$ of beam steering for a $1\,$T applied field. This behavior is enabled by the magneto-optic response of the active layer, which breaks Lorentz reciprocity, allowing the device to violate Kirchhoff's law of thermal radiation and giving rise to strongly asymmetric, directional thermal emission.

Unlike conventional directional thermal emitters, which remain symmetric about the surface normal, the asymmetry of this device enables the confinement of high emissivity to only positive or negative angles. Additionally, because the steering mechanism is magnetic rather than mechanical, the direction of peak thermal emission can in principle be modulated, enabling beam scanning using an incandescent source. The operation of this device in the $20$--$25\,\mu$m wavelength range, where coherent sources are scarce and existing options are bulky or cryogenic, highlights its potential as a compact and tunable source of directed far-infrared radiation. We emphasize that such a source is not a substitute for a laser since it is an incandescent emitter, and thus is spatially and temporally incoherent. Ultimately, the ability to dynamically control both the direction and asymmetry of thermal emission holds the potential to enable new capabilities for infrared sensing, imaging, free-space optical communication, and radiative energy technologies.

\begin{acknowledgments}
This material was based upon work supported by the National Science Foundation under grant no. ECCS-2146577. D.C. acknowledges support of the Department of Defense National Defense Science and Engineering Graduate Fellowship.
\end{acknowledgments}

\section{Methods}
We used the Electromagnetic Waves, Frequency Domain module in COMSOL Multiphysics to perform two-dimensional, full-wave, finite-element simulations of the device. We applied Floquet boundary conditions in the $\hat{x}$ direction. From the simulation, we calculated the reflectivity of incident TM-polarized light, which was then used to determine the absorptivity ($\alpha(\lambda,\theta)=1-R(\lambda,\theta)$, assuming no transmission) and emissivity ($\epsilon(\lambda,\theta)=\alpha(\lambda,-\theta)$, which is valid for this geometry). The carrier concentration, effective electron mass, and scattering rate of each layer are summarized in Table~\ref{tbl1}. The material parameters for the active layer were obtained from Ref.~\cite{Zhu_2014}, and the refractive index model for InGaAs is from Ref.~\cite{Pelouard_2022}.

\begin{table}[ht]
\centering
\caption{Material parameters used in simulations. $m_0$ is the free-electron mass.}
\setlength{\tabcolsep}{10pt}
\begin{tabular}{lccc}
\hline
\rule{0pt}{12pt}
Layer & Carrier concentration (cm$^{-3}$) & $m_e/m_0$ & $\gamma$ (Hz) \\
\hline
\rule{0pt}{12pt}
InAs$_{\mathrm{active}}$ & $7.8\times10^{17}$ & 0.033 & $1.55\times10^{11}$ \\
InAs$_{\mathrm{reflector}}$ & $5\times10^{19}$ & 0.1413 & $3.704\times10^{13}$ \\
\hline
\end{tabular}
\label{tbl1}
\end{table}

\bibliography{bibliography}

@article{Davids_2020, title={Electrical power generation from moderate-temperature radiative thermal sources}, volume={367}, DOI={10.1126/science.aba2089}, abstractNote={Moderate-temperature thermal sources (100° to 400°C) that radiate waste heat are often the by-product of mechanical work, chemical or nuclear reactions, or information processing. We demonstrate conversion of thermal radiation into electrical power using a bipolar grating-coupled complementary metal-oxide-silicon (CMOS) tunnel diode. A two-step photon-assisted tunneling charge pumping mechanism results in separation of charge carriers in pn-junction wells leading to a large open-circuit voltage developed across a load. Electrical power generation from a broadband blackbody thermal source has been experimentally demonstrated with converted power densities of 27 to 61 microwatts per square centimeter for thermal sources between 250° and 400°C. Scalable, efficient conversion of radiated waste heat into electrical power can be used to reduce energy consumption or to power electronics and sensors.}, number={6484}, journal={Science}, publisher={American Association for the Advancement of Science}, author={Davids, Paul S. and Kirsch, Jared and Starbuck, Andrew and Jarecki, Robert and Shank, Joshua and Peters, David}, year={2020}, month=3, pages={1341–1345} }

@article{Dong_2019, title={Electrical modulation of degenerate semiconductor plasmonic interfaces}, volume={126}, ISSN={0021-8979}, DOI={10.1063/1.5108905}, abstractNote={We demonstrate electrical modulation of plasmonic interfaces in semiconductor p-n++ junctions fabricated from both III–V and Si materials. Junction diodes are grown/fabricated, consisting of degenerately doped n-type material and heavily doped p-type material, where the n++ semiconductor acts as a plasmonic material capable of supporting infrared propagating surface plasmon polaritons. Devices were characterized electrically and optically, and we achieved tuning of the reflectivity under applied bias with amplitude reaching 1.5% in mid-IR wavelengths. We developed a model of electrical carrier injection at the degenerately doped interface, which we used to model the bias-dependent optical properties of the system. A strong agreement between our model and experimental results is demonstrated. The presented devices offer the opportunity for electrical modulation of propagating plasmonic modes in an all-semiconductor system.}, number={4}, journal={Journal of Applied Physics}, author={Dong, Zuoming and Vinnakota, Raj K. and Briggs, Andrew F. and Nordin, Leland and Bank, Seth R. and Genov, Dentcho A. and Wasserman, Daniel}, year={2019}, month=7, pages={043101} }

@article{Hwang_2023, title={Simultaneous Control of Spectral And Directional Emissivity with Gradient Epsilon-Near-Zero InAs Photonic Structures}, volume={35}, rights={© 2023 The Authors. Advanced Materials published by Wiley-VCH GmbH}, ISSN={1521-4095}, DOI={10.1002/adma.202302956}, abstractNote={Controlling both the spectral bandwidth and directionality of emitted thermal radiation is a fundamental challenge in contemporary photonics. Recent work has shown that materials with a spatial gradient in the frequency range of their epsilon-near-zero (ENZ) response can support broad spectrum directionality in their emissivity, enabling high total radiance to specific angles of incidence. However, this capability is limited spectrally and directionally by the availability of materials with phonon-polariton resonances over long-wave infrared wavelengths. Here, an approach is designed and experimentally demonstrated using doped III–V semiconductors that can simultaneously tailor spectral peak, bandwidth, and directionality of infrared emissivity. InAs-based gradient ENZ photonic structures that exhibit broadband directional emission with varying spectral bandwidths and directional ranges as a function of their doping concentration profile and thickness are epitaxially grown and characterized. Due to its easy-to-fabricate geometry, it is believed that this approach provides a versatile photonic platform to dynamically control broadband spectral and directional emissivity for a range of emerging applications in heat transfer and infrared sensing.}, number={39}, journal={Advanced Materials}, author={Hwang, Jae S. and Xu, Jin and Raman, Aaswath P.}, year={2023}, pages={2302956}, language={en} }

@article{Hwang_2025, title={Thermally Tunable Angular Selectivity of Broadband Directional Thermal Emission}, volume={25}, ISSN={1530-6984}, DOI={10.1021/acs.nanolett.5c02197}, abstractNote={Tuning the spatial extent of directional thermal emission across arbitrary spectral bandwidths is a challenging but critically important capability for a range of emerging applications such as thermophotovoltaics, thermal imaging, and radiative cooling. However, previous experimental demonstrations have been limited to narrow bandwidths, with the wavelengths of high emissivity themselves changing significantly as the directional response is reconfigured. Here, we present a thermally tunable photonic platform using InAs-based gradient epsilon-near-zero (ENZ) materials, functioning as a broadband directional thermal emitter whose angular extent of “thermal beaming”, and thus its angular selectivity, can be gradually reconfigured while maintaining the spectral response over a prescribed emissivity bandwidth via thermal free-carrier effects. We experimentally demonstrate two emitters achieving an 8° and 14° increase in the angular extent of their directional emissivity in the p-polarization across a fixed, broad wavelength range of operation (12.5 to 15 μm), for moderate temperatures below 450 K.}, number={19}, journal={Nano Letters}, publisher={American Chemical Society}, author={Hwang, Jae S. and Xu, Jin and Raman, Aaswath P.}, year={2025}, month=5, pages={8064–8071} }

@article{Liu_2011, title={A graphene-based broadband optical modulator}, volume={474}, rights={http://www.springer.com/tdm}, ISSN={0028-0836, 1476-4687}, DOI={10.1038/nature10067}, number={7349}, journal={Nature}, author={Liu, Ming and Yin, Xiaobo and Ulin-Avila, Erick and Geng, Baisong and Zentgraf, Thomas and Ju, Long and Wang, Feng and Zhang, Xiang}, year={2011}, month=6, pages={64–67}, language={en} }

@article{Park_2018, title={Dynamic thermal emission control with InAs-based plasmonic metasurfaces}, volume={4}, ISSN={2375-2548}, DOI={10.1126/sciadv.aat3163}, abstractNote={Scientists and engineers revealed a way to electrically control heat transfer on demand in a nano-optical device.
          , 
            Thermal emission from objects tends to be spectrally broadband, unpolarized, and temporally invariant. These common notions are now challenged with the emergence of new nanophotonic structures and concepts that afford on-demand, active manipulation of the thermal emission process. This opens a myriad of new applications in chemistry, health care, thermal management, imaging, sensing, and spectroscopy. Here, we theoretically propose and experimentally demonstrate a new approach to actively tailor thermal emission with a reflective, plasmonic metasurface in which the active material and reflector element are epitaxially grown, high-carrier-mobility InAs layers. Electrical gating induces changes in the charge carrier density of the active InAs layer that are translated into large changes in the optical absorption and thermal emission from metasurface. We demonstrate polarization-dependent and electrically controlled emissivity changes of 3.6%P (6.5% in relative scale) in the mid-infrared spectral range.}, number={12}, journal={Science Advances}, author={Park, Junghyun and Kang, Ju-Hyung and Liu, Xiaoge and Maddox, Scott J. and Tang, Kechao and McIntyre, Paul C. and Bank, Seth R. and Brongersma, Mark L.}, year={2018}, month=12, pages={eaat3163}, language={en} }

@article{Raman_2014, title={Passive radiative cooling below ambient air temperature under direct sunlight}, volume={515}, ISSN={0028-0836, 1476-4687}, DOI={10.1038/nature13883}, number={7528}, journal={Nature}, author={Raman, Aaswath P. and Anoma, Marc Abou and Zhu, Linxiao and Rephaeli, Eden and Fan, Shanhui}, year={2014}, month=11, pages={540–544}, language={en} }

@article{Sakr_2017, title={Thermophotovoltaics with spectral and angular selective doped-oxide thermal emitters}, volume={25}, ISSN={1094-4087}, DOI={10.1364/OE.25.00A880}, abstractNote={Deliberate control of thermal emission properties using nanophotonics has improved a number of applications including thermophotovoltaics (TPV), radiative cooling and infrared spectroscopy. In this work, we study the effect of simultaneous control of angular and spectral properties of thermal emitters on the efficiencies of TPV systems. While spectral selectivity reduces sub-bandgap losses, angular selectivity is expected to enhance view factors at larger separation distances and hence to provide flexibilities in cooling the photovoltaic converter. We propose a design of an angular and spectral selective thermal emitter based on waveguide perfect absorption phenomena in epsilon-near-zero thin-films. Aluminum-doped Zinc-Oxide is used as an epsilon-near-zero material with a cross-over frequency in the near-infrared. A high contrast grating is designed to restrict the emission in a range of angles around the normal direction, while an integrated filter ensures spectral selectivity to reduce sub-bandgap losses. Theoretical analysis shows an expected relative enhancement of the TPV system efficiency of at least 32% using selective emitters with ideal angular and spectral selectivity at large separation distances compared to a blackbody. This enhancement factor, however, reduces to 3.9% with non-ideal selective emitters. This big reduction of the efficiency is attributed to sub-bandgap losses, off-angular losses and hightemperature dependence of optical constants.}, number={20}, journal={Optics Express}, author={Sakr, Enas and Bermel, Peter}, year={2017}, month=10, pages={A880}, language={en} }

@article{Suhendri_2020, title={Implementation of Passive Radiative Cooling Technology in Buildings: A Review}, volume={10}, ISSN={2075-5309}, DOI={10.3390/buildings10120215}, abstractNote={Radiative cooling (RC) is attracting more interest from building engineers and architects. Using the sky as the heat sink, a radiative cooling material can be passively cooled by emitting heat to the sky. As a result of the development of material technology, RC research has been revived, with the aim of increasing the materials’ cooling power as well as ﬁnding reliable ways to utilize it in cooling for buildings. This review identiﬁes some issues in the current implementation of RC technologies in buildings from an architectural point of view. Besides the technical performance of the RC technologies, some architectural aspects, such as integration with architectural features, aesthetic requirements, as well as fully passive implementations of RC, also need to be considered for building application. In addition, performance evaluation of a building-integrated RC system should begin to account for its beneﬁt to the occupant’s health and comfort alongside the technical performance. In conclusion, this review on RC implementation in buildings provides a meaningful discussion in regard to the direction of the research.}, number={12}, journal={Buildings}, author={Suhendri and Hu, Mingke and Su, Yuehong and Darkwa, Jo and Riffat, Saffa}, year={2020}, month=11, pages={215}, language={en} }

@article{Vassant_2013, title={Electrical modulation of emissivity}, volume={102}, ISSN={0003-6951, 1077-3118}, DOI={10.1063/1.4793650}, abstractNote={We demonstrate that it is possible to modulate the thermal emission through an electrical modulation of the emissivity. The basic idea is to design a device where absorption is due to a resonant phenomenon. If the resonance can be electrically controlled, then absorption and, therefore, thermal emission can be controlled. We demonstrate this general concept using THz resonant absorption by surface phonon polaritons coupled through a gold grating. In our device, absorption is mostly due to a surface phonon mode confined in a single quantum well (QW). The THz emissivity is electrically controlled by controlling the quantum well refractive index using a gate voltage to modulate the electron density. This paves the way to electrically modulated incandescent sources.}, number={8}, journal={Applied Physics Letters}, author={Vassant, S. and Moldovan Doyen, I. and Marquier, F. and Pardo, F. and Gennser, U. and Cavanna, A. and Pelouard, J. L. and Greffet, J. J.}, year={2013}, month=2, pages={081125}, language={en} }

@article{Wang_2017, title={Mid-infrared broadband absorber of full semiconductor epi-layers}, volume={381}, ISSN={03759601}, DOI={10.1016/j.physleta.2017.02.021}, number={16}, journal={Physics Letters A}, author={Wang, Shaohua and Wang, Yufei and Zhang, Siriguleng and Zheng, Wanhua}, year={2017}, month=4, pages={1439–1444}, language={en} }

@article{Hulley_2016, title={High spatial resolution imaging of methane and other trace gases with the airborne Hyperspectral Thermal Emission Spectrometer (HyTES)}, volume={9}, rights={https://creativecommons.org/licenses/by/3.0/}, ISSN={1867-8548}, DOI={10.5194/amt-9-2393-2016}, abstractNote={Abstract. Currently large uncertainties exist associated with the attribution and quantification of fugitive emissions of criteria pollutants and greenhouse gases such as methane across large regions and key economic sectors. In this study, data from the airborne Hyperspectral Thermal Emission Spectrometer (HyTES) have been used to develop robust and reliable techniques for the detection and wide-area mapping of emission plumes of methane and other atmospheric trace gas species over challenging and diverse environmental conditions with high spatial resolution that permits direct attribution to sources. HyTES is a pushbroom imaging spectrometer with high spectral resolution (256 bands from 7.5 to 12 µm), wide swath (1–2 km), and high spatial resolution (∼ 2 m at 1 km altitude) that incorporates new thermal infrared (TIR) remote sensing technologies. In this study we introduce a hybrid clutter matched filter (CMF) and plume dilation algorithm applied to HyTES observations to efficiently detect and characterize the spatial structures of individual plumes of CH4, H2S, NH3, NO2, and SO2 emitters. The sensitivity and field of regard of HyTES allows rapid and frequent airborne surveys of large areas including facilities not readily accessible from the surface. The HyTES CMF algorithm produces plume intensity images of methane and other gases from strong emission sources. The combination of high spatial resolution and multi-species imaging capability provides source attribution in complex environments. The CMF-based detection of strong emission sources over large areas is a fast and powerful tool needed to focus on more computationally intensive retrieval algorithms to quantify emissions with error estimates, and is useful for expediting mitigation efforts and addressing critical science questions.}, number={5}, journal={Atmospheric Measurement Techniques}, author={Hulley, Glynn C. and Duren, Riley M. and Hopkins, Francesca M. and Hook, Simon J. and Vance, Nick and Guillevic, Pierre and Johnson, William R. and Eng, Bjorn T. and Mihaly, Jonathan M. and Jovanovic, Veljko M. and Chazanoff, Seth L. and Staniszewski, Zak K. and Kuai, Le and Worden, John and Frankenberg, Christian and Rivera, Gerardo and Aubrey, Andrew D. and Miller, Charles E. and Malakar, Nabin K. and Sánchez Tomás, Juan M. and Holmes, Kendall T.}, year={2016}, month=6, pages={2393–2408}, language={en} }

@article{Lucchi_2018, title={Applications of the infrared thermography in the energy audit of buildings: A review}, volume={82}, ISSN={13640321}, DOI={10.1016/j.rser.2017.10.031}, abstractNote={From a wide range of bibliography (148 publications composed by books, guidelines, scientiﬁc papers, and other documents), the study presents a critical review of the use of the infrared thermography (IRT) survey in the building energy audit. After explaining its historical growth, the applicability of passive and active approaches has been described, considering well-established and emerging techniques, general procedures, types of IRcamera used, technical issues, and limitations. The passive approach is the most common to detect thermally signiﬁcant defects. Thus, a speciﬁc procedure for the energy audit has been reported, matching diﬀerent standards, guidelines, and professional advice. Similarly, recurring energy related problems are toughly presented (i.e. thermal characterization of buildings; thermal bridging, insulation level, air leakage and moisture detection; indoor temperature and U-value measurements; human comfort assessment). Finally, advantages and potential sources of errors as well as future trends in the use of IRT for the energy audit have been described. The research aims to serve as a reference for energy auditors and thermographers, to decide upon the best procedure for detecting speciﬁc energy defects.}, journal={Renewable and Sustainable Energy Reviews}, author={Lucchi, Elena}, year={2018}, month=2, pages={3077–3090}, language={en} }

@inbook{vollmer_2021,
  author    = "Michael Vollmer",
  title     = "Computer Vision: A Reference Guide",
  chapter   = "Infrared Thermal Imaging",
  publisher = "Springer International Publishing",
  year      = "2021",
  pages     = "666-670"
}

@article{Palchetti_2016, title={Remote sensing of cirrus cloud microphysical properties using spectral measurements over the full range of their thermal emission}, volume={121}, rights={http://onlinelibrary.wiley.com/termsAndConditions#vor}, ISSN={2169-897X, 2169-8996}, url={https://agupubs.onlinelibrary.wiley.com/doi/10.1002/2016JD025162}, DOI={10.1002/2016JD025162}, abstractNote={The thermal emission of cirrus clouds, spectrally resolved in the 100–1400 cm−1 range (100–7.1 μm), has been modeled and compared with measurements performed during two ﬁeld campaigns from the ground-based site of Testa Grigia on the Italian Alps at 3480 m of altitude. The analysis of cirrus microphysics, through spectral ﬁtting, shows the importance of using also the far infrared portion of the emitted spectrum at wave numbers below the 667 cm−1 carbon dioxide absorption band, where only a few measurements exist because of the high opacity of the atmosphere caused by the strong water vapor absorption. The resulted distribution of the ﬁtted cloud parameters is in good agreement with the typical statistical distribution of the midlatitude cirrus cloud parameters.}, number={18}, journal={Journal of Geophysical Research: Atmospheres}, author={Palchetti, L. and Di Natale, G. and Bianchini, G.}, year={2016}, month=9, language={en} }

@article{Shayegan_2023, title={Direct observation of the violation of Kirchhoff’s law of thermal radiation}, volume={17}, rights={2023 The Author(s), under exclusive licence to Springer Nature Limited}, ISSN={1749-4893}, DOI={10.1038/s41566-023-01261-6}, abstractNote={Thermal emission—the process through which all objects with a finite temperature radiate electromagnetic energy—has generally been thought to obey reciprocity, where the absorbed and emitted radiation from a body are equal for a given wavelength and angular channel. This equality, formalized by Gustav Kirchhoff in 1860, is known as Kirchhoff’s law of thermal radiation and has long guided designs to control the emitted radiation. Removing the constraint of Kirchhoff’s law unlocks a multitude of applications and designs for thermal emitters. Decoupling the absorptivity and emissivity relationship can be leveraged to achieve novel functions, ranging from reducing re-emission losses to the Sun in the context of solar energy harvesting systems to radiative camouflage. Here we report the direct measurements of an inequality between the spectral directional emissivity and absorptivity for a photonic design that supports a guided-mode resonance coupled to a magneto-optic material. This inequality occurs under the application of an in-plane magnetic field that modifies the normally diagonal permittivity tensor to a non-diagonal tensor in magneto-optic InAs, resulting in an antisymmetric relationship where the magnetic tuning of enhanced emissivity for a given angle correlates with decreased absorptivity for the same angle.}, number={10}, journal={Nature Photonics}, publisher={Nature Publishing Group}, author={Shayegan, Komron J. and Biswas, Souvik and Zhao, Bo and Fan, Shanhui and Atwater, Harry A.}, year={2023}, month=10, pages={891–896}, language={en} }

@article{Shayegan_2022, title={Nonreciprocal infrared absorption via resonant magneto-optical coupling to InAs}, volume={8}, DOI={10.1126/sciadv.abm4308}, abstractNote={Nonreciprocal elements are a vital building block of electrical and optical systems. In the infrared regime, there is a particular interest in structures that break reciprocity because their thermal absorptive (and emissive) properties should not obey the Kirchhoff thermal radiation law. In this work, we break time-reversal symmetry and reciprocity in n-type–doped magneto-optic InAs with a static magnetic field where light coupling is mediated by a guided-mode resonator structure, whose resonant frequency coincides with the epsilon-near-zero resonance of the doped indium arsenide. Using this structure, we observe the nonreciprocal absorptive behavior as a function of magnetic field and scattering angle in the infrared. Accounting for resonant and nonresonant optical scattering, we reliably model experimental results that break reciprocal absorption relations in the infrared. The ability to design these nonreciprocal absorbers opens an avenue to explore devices with unequal absorptivity and emissivity in specific channels.}, number={18}, journal={Science Advances}, publisher={American Association for the Advancement of Science}, author={Shayegan, Komron J. and Zhao, Bo and Kim, Yonghwi and Fan, Shanhui and Atwater, Harry A.}, year={2022}, month=5, pages={eabm4308} }

@article{Wilson_2019, title={An intense, few-cycle source in the long-wave infrared}, volume={9}, rights={2019 The Author(s)}, ISSN={2045-2322}, DOI={10.1038/s41598-019-42433-1}, abstractNote={For the last several decades, the wavelength range accessible for strong-field, few-cycle studies has remained limited to the visible, near infrared and mid-wave infrared regimes. In particular, sources in the long-wave infrared have been lacking. We report the development of a 1 kHz, few-cycle laser source with up to a 9 μm central wavelength and gigawatt peak powers. When focused, this source can ionize gas targets, which we demonstrate here through the ionization of atomic xenon at wavelengths ranging from 5 μm to 9 μm. This opens up new opportunities for fundamental atomic and molecular physics, enabling experimental tests of strong-field ionization theories in the extreme long-wavelength, few-cycle limit and the direct excitation of vibrational transitions in organic molecules.}, number={1}, journal={Scientific Reports}, publisher={Nature Publishing Group}, author={Wilson, Derrek J. and Summers, Adam M. and Zigo, Stefan and Davis, Brandin and Robatjazi, Seyyed-Javad and Powell, Jeffery A. and Rolles, Daniel and Rudenko, Artem and Trallero-Herrero, Carlos A.}, year={2019}, month=4, pages={6002}, language={en} }

@article{Cui_2026, title={Nonreciprocal perturbation theory for near-complete violation of thermal emission}, abstractNote={Nonreciprocal thermal emitters that violate Kirchhoff’s law of thermal radiation hold promise for energy harvesting, sensing, and imaging, yet our understanding of such photonic systems has largely relied on phenomenological models or brute-force numerical optimization. Here we develop a dispersive perturbation theory for magneto-optical plasmonic semiconductors that yields a closed-form analytical expression for the resonance frequency shift induced by an applied magnetic field. The expression reveals that this shift is governed by the spatial overlap of the optical spin angular momentum density of the resonant mode with the magneto-optical material. We validate the theory against full-wave finite-element simulations across a range of incidence angles and field strengths, and use it to design an InAs/InGaAs metasurface that achieves a nonreciprocal emissivity vs. absorptivity contrast exceeding 0.8 at only 0.05 T applied magnetic field. By comparing two structurally distinct guided-mode-resonance designs, we show that the theory quantitatively accounts for a two-order-of-magnitude difference in magnetic field sensitivity, tracing it to differences in mode confinement, spin density overlap, and carrier effective mass. These results establish a first-principles framework for engineering nonreciprocal thermal emitters with maximal Kirchhoff’s law violation at practically accessible magnetic field strengths.}, author={Cui, Daniel and Raman, Aaswath P}, language={en} ,year={2026}, month=3, journal={arxiv}}

@article{Lin_2003, title={Three-dimensional photonic-crystal emitter for thermal photovoltaic power generation}, volume={83}, ISSN={0003-6951, 1077-3118}, DOI={10.1063/1.1592614}, abstractNote={A three-dimensional tungsten photonic crystal is experimentally realized with a complete photonic band gap at wavelengths λ⩾3 μm. At an effective temperature of 〈T〉∼1535 K, the photonic crystal exhibits a sharp emission at λ∼1.5 μm and is promising for thermal photovoltaic (TPV) power generation. Based on the spectral radiance, a proper length scaling and a planar TPV model calculation, an optical-to-electric conversion efficiency of ∼34% and electrical power of ∼14 W/cm2 is theoretically possible.}, number={2}, journal={Applied Physics Letters}, author={Lin, S. Y. and Moreno, J. and Fleming, J. G.}, year={2003}, month=7, pages={380–382}, language={en} }
\end{document}